\documentclass[superscriptaddress,reprint,citeautoscript,prb]{revtex4-1}
\usepackage{graphicx} 
\usepackage[margin=1in]{geometry} 
\usepackage{amsmath}
\usepackage{hyperref} 
\usepackage{multirow}
\usepackage{newtxtext}                                        
\usepackage{booktabs}
\usepackage{xcolor} 
\usepackage{hyperref}
\usepackage{mathtools}
\usepackage[normalem]{ulem}
\hypersetup{colorlinks, 
	linkcolor={blue!75!black!80!yellow},
	citecolor={blue!75!black!80!yellow}, 
	urlcolor={blue!75!black!80!yellow}
	}
\usepackage[cmintegrals]{newtxmath}
\makeatletter \renewcommand\@make@capt@title[2]{%
\@ifx@empty\float@link{\@firstofone}{\expandafter\href\expandafter{\float@link}}%
\sffamily{\textbf{#1}}\@caption@fignum@sep#2 }

\usepackage{xcolor}
\usepackage[textwidth=1.5cm,textsize=footnotesize]{todonotes}

\newcommand{\sub}[1]{\ensuremath{_{\textrm{#1}}}} 
\newcommand{\HarvardSEAS}{John A. Paulson School of Engineering and Applied
Sciences, Harvard University, Cambridge, MA, USA}
\newcommand{\HarvardCCB}{Department of Chemistry and Chemical Biology, Harvard
University, Cambridge, MA, USA}

\begin{document}

\preprint{AIP/123-QED}

\title{Dynamic modulation of phonon-assisted transitions
in quantum defects\\ in monolayer transition-metal dichalcogenide semiconductors}

\author{Chitraleema Chakraborty}
\thanks{These authors contributed equally}
\affiliation{\HarvardSEAS}
\author{Christopher J. Ciccarino}
\thanks{These authors contributed equally}
\affiliation{\HarvardSEAS}
\affiliation{\HarvardCCB}
\author{Prineha Narang}
\email{prineha@seas.harvard.edu}
\affiliation{\HarvardSEAS}
\date{\today}

\begin{abstract}
\noindent Quantum localization \emph{via} atomic point defects in semiconductors is of significant fundamental and technological importance. Quantum defects in monolayer transition-metal dichalcogenide semiconductors have been proposed as stable and scalable optically-addressable spin qubits. Yet, the impact of strong spin-orbit coupling on their dynamical response, for example under optical excitation, has remained elusive. In this context, we study the effect of spin-orbit coupling on the electron-phonon interaction in a single chalcogen vacancy defect in monolayer transition metal dichalcogenides, molybdenum disulfide (MoS\sub{2}) and tungsten disulfide (WS\sub{2}). From \emph{ab initio} electronic structure theory calculations,  we find that spin-orbit interactions tune the magnitude of the electron-phonon coupling in both optical and charge-state transitions of the defect, modulating their respective efficiencies. This observation opens up a promising scheme of dynamically modulating material properties to tune the local behavior of a quantum defect. 
\end{abstract}

\maketitle

Defects are ubiquitous in materials. At low enough densities, point defects within a lattice can be treated as effectively isolated from each other. These defects create broken bonds which can lead to localized electronic orbitals, with behavior similar to trapped atoms or molecules, and can therefore be used as qubits, building blocks for various applications in the field of quantum information science, ranging from quantum computing, communications to nanoscale sensing~\cite{ narang_quantum_2019,bassett_quantum_2019,awschalom_quantum_2018,aharonovich_solid-state_2016, neuman_selective_2020}. Prominent examples of optically active quantum defects include color centers in Diamond, Silicon Carbide (SiC), hexagonal Boron Nitride (h-BN), etc.~\cite{weber_quantum_2010,doherty_nitrogen-vacancy_2013, tran_quantum_2016} Some of these exhibit sufficiently long quantum coherence, which makes them suitable for small-scale quantum information systems~\cite{atature_material_2018}. 
However, no single solid-state candidate overcomes the challenges of scalability and photonic integration, to enable large scale quantum information and communication protocols, in addition to  atomically addressable quantum operations~\cite{atature_material_2018}.

Two-dimensional materials are a promising platform for scalable quantum technologies~\cite{chakraborty_advances_2019}. Their low dimensionality and versatility of growth of monolayers over various substrates\cite{briggs_roadmap_2019} makes them easy to integrate in various quantum architectures. Single-photon emitters have been found in various 2D materials such as transition metal dichalcogenides (TMDCs)~\cite{chakraborty_voltage-controlled_2015, srivastava_optically_2015, he_single_2015, koperski_single_2015}, transition metal mono-chalcogenides~\cite{tonndorf_single-photon_2017} and h-BN~\cite{tran_quantum_2016}.
These systems have, so far, demonstrated promising properties including quantum
emission up to room temperature~\cite{tan_unraveling_2019,tran_quantum_2016, luo_single_2019},
electrostatic control of spin-valley-photon interfaces in TMDCs~\cite{lu_optical_2019,chakraborty_3d_2018,chakraborty_quantum-confined_2017},
and coherence times up to microseconds~\cite{qian_defect_2020, moody_microsecond_2018}.
The possibility of deterministic positioning~\cite{branny_deterministic_2017, palacios-berraquero_large-scale_2017} and control~\cite{chakraborty_strain_2020} of optically active emitters \emph{via} strain engineering has also been demonstrated. Importantly, 2D material platforms also offer the opportunity to image~\cite{hayee_revealing_2020} and controllably fabricate  quantum defects at atomically precise locations using present day scanning probe technologies like aberration corrected transmission electron microscopy and low temperature scanning tunnelling microscopy~\cite{dyck_atom-by-atom_2019, tian_correlating_2020}. This would enable arrays of quantum defects integrated in compact quantum architectures.

Rational design of quantum defects requires careful consideration of both the defect 
and host material, which jointly determine the  behavior and properties.
Requirements of the host material include important descriptors such as the band-gap size, concentration of the nuclear spin bath, potential for high quality material growth, and spin-orbit interactions,
are at least partially determined by the host material. This has most notably been successful in the emergence of defects in silicon
carbide~\cite{weber_quantum_2010,bassett_quantum_2019}. Further, electron-phonon interactions \cite{brown_ab_2016, brown_nonradiative_2016} also play an important role in the physical properties of materials including defects. For example, a high  Debye-Waller factor for improved electronic transitions into the zero phonon line is desired for an optically active defect~\cite{bassett_quantum_2019}.
In TMDCs, previous studies have reported defect engineering in such moderately gap materials~\cite{klein_scalable_2020, zheng_point_2019} with optically active defects
in high-quality epitaxially grown materials~\cite{wu_locally_2019}. Meanwhile, the reduced dimensionality of
2D systems in general allows for reduced nuclear-spin interactions, and therefore longer spin coherence compared to 3D materials~\cite{ye_spin_2019}.
Thus far, the effect of spin-orbit (SO) coupling (SOC) on the host material has not been explored. Unlike notable 3D (Diamond and SiC) and 2D (h-BN) hosts, monolayer TMDCs can have spin-orbit splittings on the order of hundreds of meVs \cite{ciccarino_dynamics_2018}, making SOC in these host-systems a distinctive feature. Such strong spin-orbit effects has important implications for defect behavior~\cite{schuler_large_2019}.

\begin{figure*}[ht]
\includegraphics[scale=0.8]{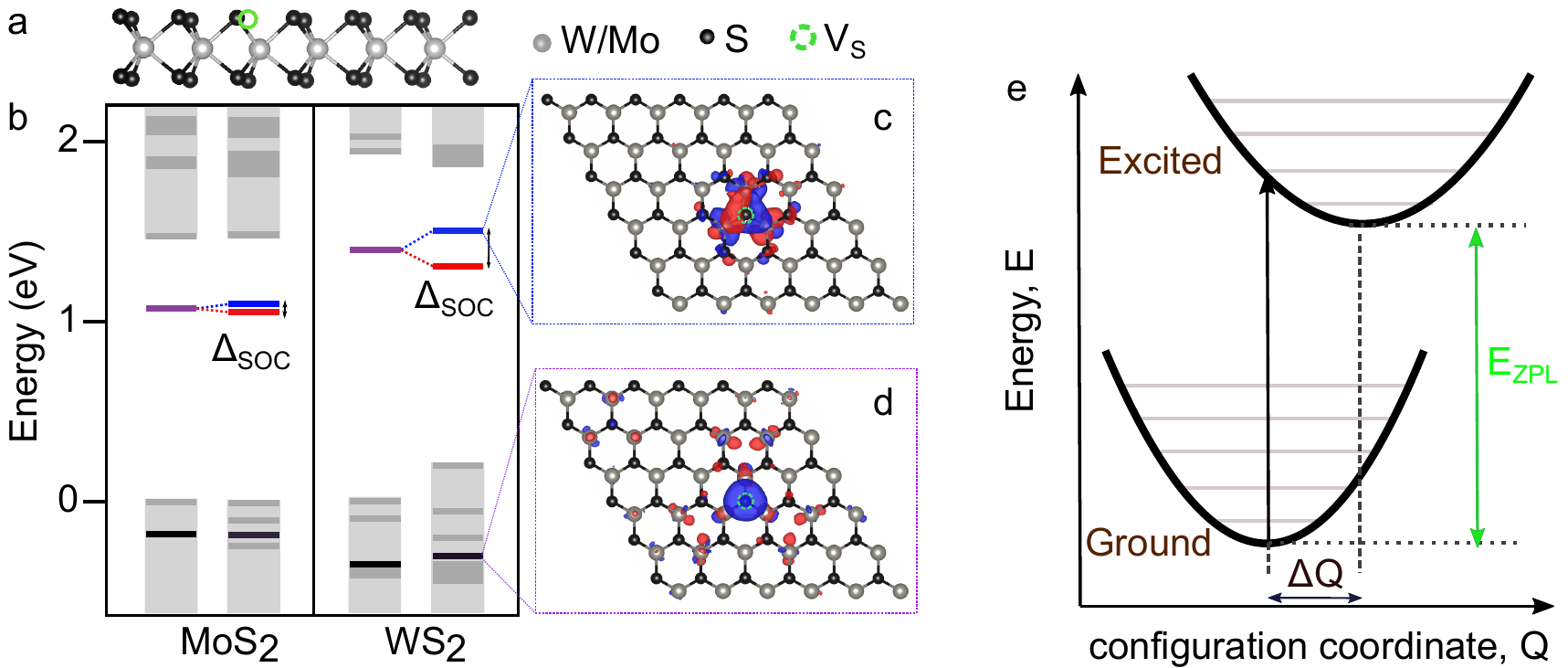}
\caption{a. Side view of the TMDC lattice with a single chalcogen defect illustrated with the green dashed circle, b. Comparison of ground state energy level diagram of the chalcogen vacancy defect excluding (left column) and including (right column) the effect of spin-orbit coupling in MoS\sub{2} and WS\sub{2}. Grey lines represent bulk states. Solid purple lines are localized defect orbitals without SOC. Red and blue solid lines represent spin-orbit split mid-gap defect orbitals. Black lines represent defect states within the valence band. c,d. Top view of the wavefunction amplitude of the defect orbital in the valence band (bottom) and the midgap orbitals (top). e. Franck-Condon diagram of the ground and excited states of the V\sub{S} defect. The vertical axis represents energy and the horizontal axis represents the effective nuclear displacement along the direction connecting the two geometries, given by $\Delta Q$. $E_{\mathrm{ZPL}}$ is the zero phonon line energy. 
}
\label{fig:1}
\end{figure*}

Here, we theoretically study the effects of spin-orbit coupling on the dynamic lattice response upon excitation of a sulfur-vacancy (V\sub{S}) defect in monolayer TMDCs MoS\sub{2} and WS\sub{2}. These defects have been shown to be experimentally accessible~\cite{schuler_large_2019,hong_exploring_2015}. Many of the unique optical properties and static electronic structure have been studied
both experimentally and theoretically~\cite{haldar_systematic_2015, tanoh_enhancing_2019}. However, the nature of the excited state transitions, and in particular the effects of phonons \cite{gupta_franck_2018}, has been largely unexplored in these systems. 
Here, we calculate the vibrational contributions to the electronic transitions in these V\sub{S}
centers in MoS\sub{2} and WS\sub{2}. Our results highlight the critical 
importance of spin-orbit interactions on the optical transitions observed in these defect systems. 
Further, these results suggest new opportunities to tune the efficiency of electronic transitions by accessing different defect states presented by the strong spin-orbit interactions.

 
\begin{figure}[ht]
\includegraphics[width=\columnwidth]{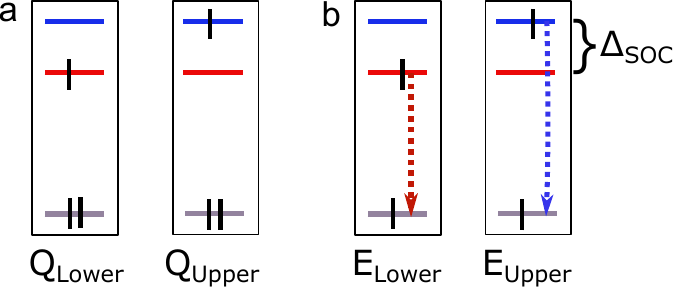}
\caption{Excited states describing a. charge capture (Q) and b. optical excitation (E) from highest occupied molecular orbital to the lower (red, Q\sub{Lower}, E\sub{Lower}) and upper (blue, Q\sub{Upper}, E\sub{Upper}) spin-orbit split mid-gap defect orbitals.}
\label{fig:2}
\end{figure}

The V\sub{S} defect we consider in monolayer MoS\sub{2} and WS\sub{2} is depicted in Fig.~\ref{fig:1}a. We perform density functional theory (DFT) calculations using the PBEsol exchange-correlation functional within JDFTx~\cite{sundararaman_jdftx_2017}. To model the isolated defect, a $6\times6\times1$ supercell of the primitive hexagonal unit cell is constructed, with a single sulfur atom removed. Spin-orbit coupling was included by use of fully relativistic pseudopotentials. We sample the Brillouin zone of the supercell using the $\Gamma$ point approximation. The phonons are computed via finite displacement of the same supercell.

Figure~\ref{fig:1}b presents the predicted level structure of the defects. Similar to previous predictions~\cite{schuler_large_2019}, we find, in addition to an occupied defect state within the valence band, the sulfur vacancy introduces two in-gap orbitals, which
are unoccupied in the ground state. 
These orbitals are energetically split only by the spin-orbit interaction.
The heavy mass of the constituent atoms of these TMDCs helps give rise to stronger spin-orbit splitting than in previously studied host materials like diamond or silicon carbide. Such splittings can be on the order of 100's of meV (depending on the TMDC) and strongly influences the optical and electrical properties of these monolayers~\cite{kosmider_large_2013}. As presented in the energy level diagram in Fig.~\ref{fig:1}b, we observe a splitting of 46 meV in the in-gap V\sub{S} orbitals of MoS\sub{2}. We next compare this splitting with the V\sub{S} defect in WS\sub{2}, which has a stronger spin-orbit coupling than MoS\sub{2}. We observe a correspondingly larger splitting of the in-gap orbitals of 194 meV. Such large splitting of the in-gap states was also previously observed in the scanning tunnelling spectra of sulfur vacancy in WS\sub{2}~\cite{schuler_large_2019}. Meanwhile, the occupied orbital below the valence edge is $s$-like and therefore no spin-orbit splitting is observed in our calculations. 
The contribution of the unoccupied mid-gap orbitals consists primarily of $d$ orbitals of the heavy W atoms, which are responsible for the large magnitude of the spin-orbit splitting~\cite{schuler_large_2019}. Figures~\ref{fig:1}c-d presents  these specific defect wavefunctions. The wavefunctions of the V\sub{S} defect states are symmetrically localized around the sulfur vacancy for both the occupied and the unoccupied states.

\begin{figure*}[ht]
\includegraphics[scale=0.75]{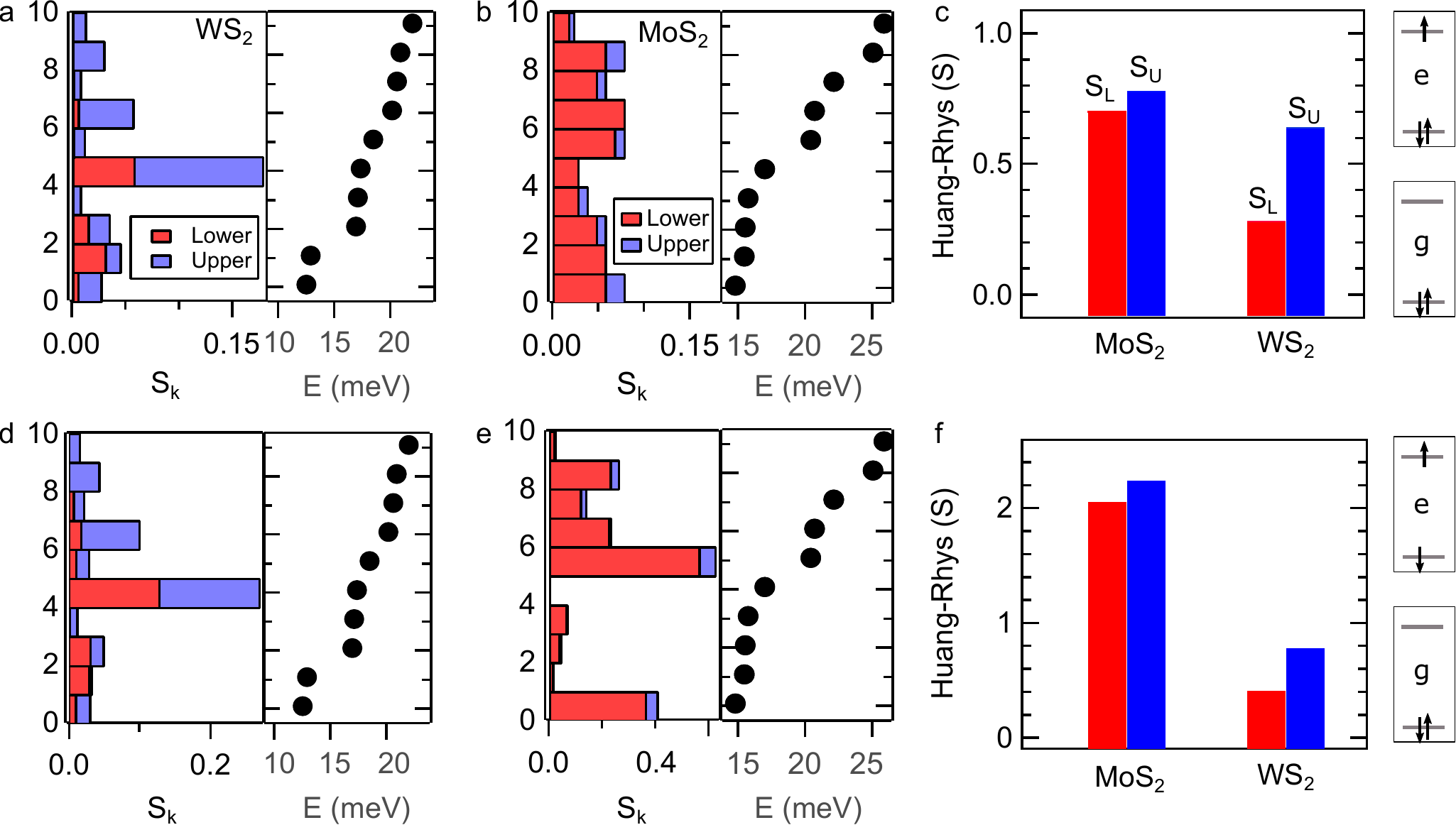}
\caption{The ten highest contributing vibrational modes to the charge capture (optical excitation) process for a(d) WS\sub{2} and b(e) MoS\sub{2}. Red (blue) bars indicate the partial Huang-Rhys factor, $S_k$, when the lower (higher) energy spin-orbit split defect orbital is occupied with a single electron. Panels c. and f. represent the total 
Huang-Rhys factor for a charge capture and optical excitation process, respectively. The
blue and red bars distinguish transitions involving the lower (S\sub{L}) and
upper (S\sub{U}) spin-orbit split defect orbitals. Insets of c. and f. depicts electronic configurations of the charge capture and optical excitation processes, respectively.}
\label{fig:3}
\end{figure*}

We study the phonon coupling parameters for two types of excited states as described in Fig.~\ref{fig:2}. The first excited state considered here is the singly charged state of the defect (Fig.~\ref{fig:2}a) where an extra electron is added and occupies one of the  unfilled defect orbitals within the gap. The second excited state,
shown in Fig.~\ref{fig:2}b, depicts an electron promoted to the unfilled defect orbital from the lowest occupied energy level (i.e., the valence band edge).
In both cases, the electronic excitations are instantaneous processes so the transition probability can be calculated at a fixed nuclear position. This allows us to utilize the Franck-Condon principle.\cite{alkauskas_tutorial_2016}
In this picture, the instantaneous electronic transition is followed by nuclear motion, such that the atomic nuclei rearrange between the ground and excited electronic configurations, which in turn can induce crystal vibrations (i.e., phonons). The transition process can be illustrated using potential energy diagrams, such as that shown in Fig.~\ref{fig:1}e.
Here, the horizontal axis denotes the nuclear configuration coordinate of the combined defect and host
system. After the electronic transition, the configuration of the atoms in the lattice surrounding the defect will change, which is seen by the horizontal offset between the two parabolas, $\Delta Q$. This can be formally defined as the mass-weighted effective displacement of the atoms relaxed in the two electronic 
configurations~\cite{alkauskas_tutorial_2016}:
\begin{equation}
    \Delta Q = \sum_{\alpha, i} m_{\alpha}^{1/2} 
    (R_{e,\alpha i}-R_{g,\alpha i}).
\label{eq:deltaQ}
\end{equation}
where $m_\alpha$ is the mass of atom $\alpha$, $R$ is the corresponding relaxed position for the ground ($g$) and excited ($e$) electronic configurations, with $i$ indexing the Cartesian direction. Returning to Fig.~\ref{fig:1}e, 
absorption of an electron is represented by a vertical arrow. At this point, the defect enters a non-equilibrium nuclear configuration and will relax into a lower-energy vibrational state. This relaxation process can involve emission of phonons.
A typical fluorescence or an electron tunnelling spectroscopy therefore consists of phonon sidebands in addition to the E$_{\mathrm{ZPL}}$ peak that results from transitions between the lowest vibration state of the ground electronic level and that of the excited state.

To study the vibrational effects of the transitions described in Fig.~\ref{fig:2}, 
we either add an additional charge to the defect supercell system (Fig.~\ref{fig:2}a, charge capture), or we constrain the occupation of the electronic occupation to mimic an optically-excited configuration (Fig.~\ref{fig:2}b, optical excitation). The latter is 
done using constrained-DFT within JDFTx; in both excitation cases we look at the difference in geometries for the ground- and excited-state configurations ($\Delta$SCF). After relaxing the ground and excited-state 
ionic positions, we can use Eq.~\ref{eq:deltaQ} to define the collective displacement of
each transition ($\Delta Q$). These values are presented in Table~\ref{tab:1}. 
For both the charge capture and optical excitation transitions in MoS\sub{2}, we find $\Delta Q$ values which are significantly larger than in WS\sub{2}. From an intuition perspective, the larger the movement of atoms upon excitation, the more likely it is for the defect to incorporate phonons in the transition. Further, the relative displacement of atoms between excitation to upper and lower split orbitals is much higher ($> 40$\% difference) for WS\sub{2} than MoS\sub{2} ($\sim5$\% difference).


\begin{table}
\begin{tabular}{|p{3cm}|p{2cm}|p{2cm}|}
 \hline
 & MoS\sub{2}  & WS\sub{2} \\
 \hline
 $\Delta Q^{L}$, Charged & 0.59 & 0.37 \\
 $\Delta Q^{U}$, Charged & 0.62 & 0.56 \\
 $\Delta Q^{L}$, Optical & 0.92 & 0.42 \\
 $\Delta Q^{U}$, Optical & 0.96 & 0.60 \\
 $\frac{\Delta Q^U-\Delta Q^L}{\Delta Q^L}$ \%, Charged & 5\% & 47\% \\
 $\frac{\Delta Q^U-\Delta Q^L}{\Delta Q^L}$ \%, Optical & 4.7\% & 44\% \\
 \hline
\end{tabular}
\caption{Effective displacements ($\Delta Q$, defined in Eq.~\ref{eq:deltaQ}) of the supercell in units of amu$^{1/2}$ \AA.}
\label{tab:1}
\end{table}

\begin{figure}[ht]
\includegraphics[width=\columnwidth]{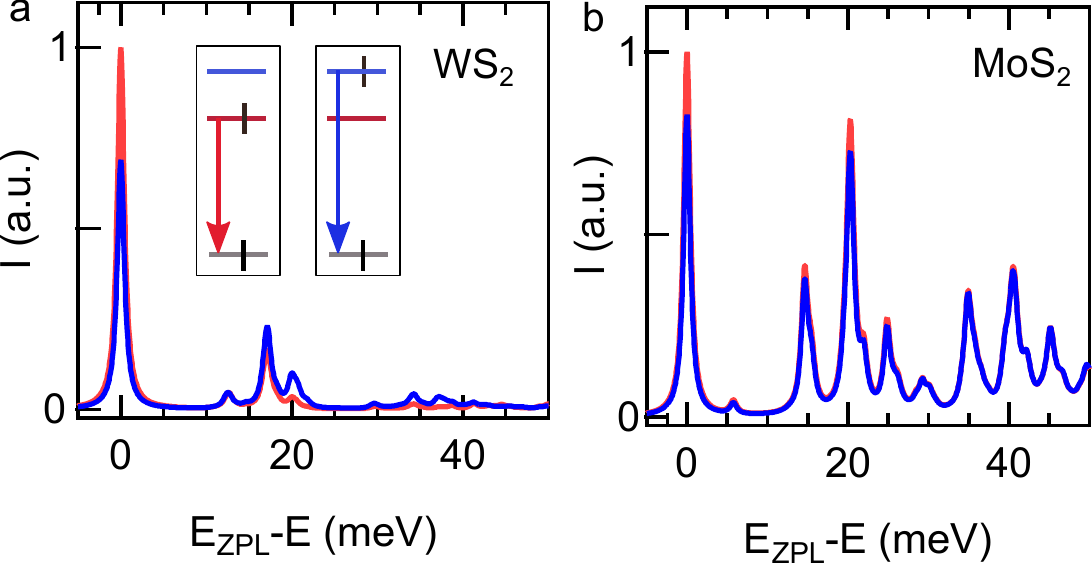}
\caption{Phonon Spectra of V\sub{S} defects in a. WS\sub{2} and b. MoS\sub{2} describing the zero phonon line and phonon sidebands. Red (Blue) curve indicates emission spectra when the lower (higher) energy spin-orbit split defect level is occupied with a single electron from the highest occupied orbital of the given system.}
\label{fig:4}
\end{figure}

We next calculate the partial Huang-Rhys factor $S_k$, which is the average number of phonons of a particular mode $k$ with energy $E_k$ emitted during the transition from ground to excited state.~\cite{alkauskas_tutorial_2016}
Our \emph{ab initio} calculations include the vacancy explicitly and therefore capture both the bulk-like and quasi-local phonon modes that can potentially couple to the transition.
The ten phonon modes with the largest $S_k$ values for each of the transitions studied are shown in Fig.~\ref{fig:3}. The results for the charge-state transition are summarized in Fig.~\ref{fig:3}a-c. Comparing the $S_k$ values for WS\sub{2}, we find that there is a significant difference when the electron is in the lower (red) or upper (blue) SO-split state. The highest contributing vibrational mode of energy 17.15 meV has a difference of 67\% in its $S_k$ value (Fig.~\ref{fig:3}a). This mode corresponds to localized motion involving the three nearest transition metal atoms neighboring the chalocgen defect. Comparing with MoS\sub{2} in Fig.~\ref{fig:3}b, the difference in the $S_k$ value for the highest contributing phonon mode (with energy 20.22 meV) is 25\%. 

The total HR factor $S$ ($= \sum_k S_k$) is the total number of phonons that on average are emitted during the transition.  
Figure~\ref{fig:3}c summarizes the total HR factor for the different transitions considered in the V\sub{S} defect for both MoS\sub{2} and WS\sub{2}. The difference in the red and blue bars represent the modulation in the HR factor due to the transitions into the respective SO-split orbital. For WS\sub{2},
with larger spin-orbit splitting, the modulation of the Huang-Rhys factor ($\Delta S$) is much larger ($\sim$55\%) than MoS\sub{2} ($\sim$10\%). 

We perform the same calculations for the optical transition depicted in the inset of Fig.~\ref{fig:3}f. In Fig.~\ref{fig:3}d-e we have reported the contributions from the same vibrational modes as Fig.~\ref{fig:3}a-b. We have also found a significant difference in $S_k$ values for the optical excitation (Fig.~\ref{fig:3}d-e) to the lower (red) as compared to the upper (blue) SO-split state, similar to the case of the charge capture process (Fig.~\ref{fig:3}a-b).
As reported in Table~\ref{tab:1}, the difference in effective displacement in the excited state was only 5\% for MoS\sub{2}. It therefore follows that the calculated contributions of the vibrational modes that were involved in the transition are similar.    
However, for WS\sub{2}, where the spin-orbit splitting is a factor of five higher than MoS\sub{2}, this effective displacement varied from 44-47\% with calculated difference of 50-55\% within the values of $S_\mathrm{L}$ and $S_\mathrm{U}$ for the two different excited state configurations. This difference in the phonon mediated transition for the lower ($S_\mathrm{L}$) and the upper ($S_\mathrm{U}$) split states directly relates to the magnitude of the spin-orbit splitting present.
From these observations, we conclude spin-orbit coupling to be the most dominant parameter that controls the efficiency of the coupling to the spin-orbit split mid-gap defect levels
for both types of transitions considered here.  
To describe the vibrational broadening of the transition spectra, we have summed up all possible transitions between the vibrational levels in the ground and those in the excited states. The normalized lineshape of the spectra can be calculated using a generating function approach as described previously~\cite{alkauskas_first-principles_2014}.
Similar sidebands were previously confirmed experimentally by scanning tunnelling spectroscopy in a chalcogen vacancy defect in monolayer WS\sub{2}~\cite{schuler_large_2019} and can be also experimentally verified via fluorescence spectroscopy in the future. 
Figure~\ref{fig:4}a-b presents the total spectra on promoting an electron from the lowest occupied level to the lower (red) or upper (blue) SO split level. 
The corresponding phonon energies for the two cases is similar as also presented in Fig.~\ref{fig:3}d-e. However, higher contributions to the ZPL is predicted for the lower (red) split state in WS\sub{2} vs MoS\sub{2} which stems from the difference in the Huang-Rhys factors reported above.

\begin{figure}[ht]
\includegraphics[scale=0.9]{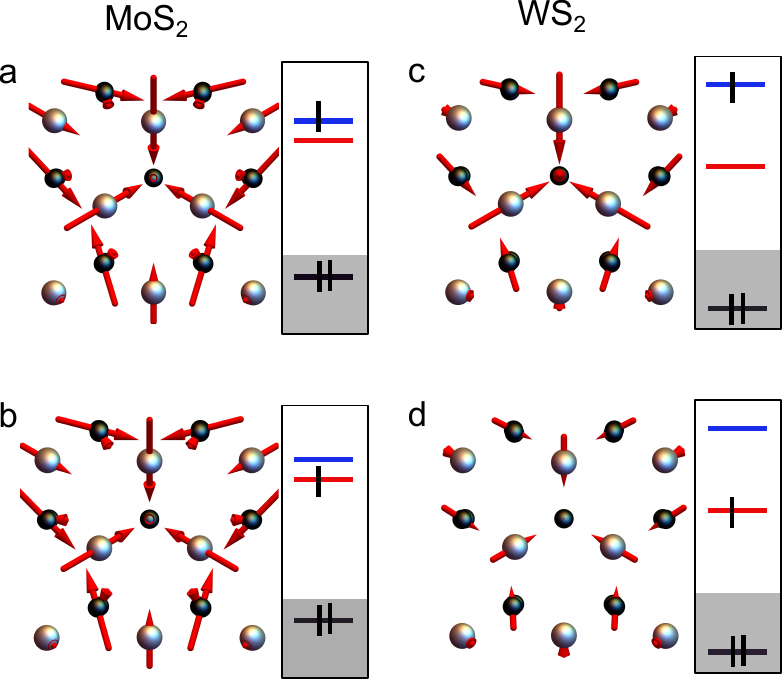}
\caption{ a-b Displacement vectors of the atoms near the V\sub{S} defect in MoS\sub{2} and c-d. WS\sub{2} for the corresponding excitations described by the energy level diagrams on the right. 
The total difference in magnitude of the effective displacement of the atoms in the defect leads to large difference in the phonon coupling in WS\sub{2} as compared to MoS\sub{2} (quantified in Table~\ref{tab:1}).}
\label{fig:5}
\end{figure}

To understand the influence of the spin-orbit splitting on the transition properties of the defect, we compared the differences in geometry shift due to electron occupation in the lower versus the upper SO-split defect orbitals. The shift in the relaxed defect cells for excitation to either of the two split levels are presented in Fig.~\ref{fig:5}. 
From the $\Delta$SCF calculations for MoS\sub{2}, significant geometry shifts occurred in the defect cell for both the excited states. However the nature of the shift of the excited state is similar irrespective of which level is being resonantly populated with an electron (Fig.~\ref{fig:5}a-b). The arrows represent the extent and direction of the shift. However, for WS\sub{2}, we observe a marked difference in magnitude of the first nearest neighboring tungsten atom as well as a significant shift in the relative direction of the first neighboring groups of sulfur atoms (Fig.~\ref{fig:5}c-d). This difference is directly related to the difference in the SO splitting for the two host materials. Such a difference in the movement of atoms upon excitation leads to a difference in the strength of phonon coupling in the transition from ground to excited state. We note that additional effects for these and other TMDC-based defects should be considered, including the role of charge state and the potential for Jahn-Teller effects~\cite{harris_group_2019,ciccarino_strong_2020}. These effects could be the subject of future work.

In conclusion, we predict a strong spin-orbit splitting in the localized mid-gap orbitals of the V\sub{S} defect in TMDC monolayers. This SO splitting alters the relaxed excited-state geometry which influences the contribution of vibrational modes to the transitions studied. Specifically, the modulation of the Huang-Rhys factor as the excitation is switched from lower to upper split state is strongly dependent on the spin-orbit splitting of a given TMDC. In future, such geometry shifts and therefore defect-phonon coupling can provide an additional fingerprint that can be used to identify defect candidates seen in scanning probe experiments.
Looking ahead, this also provides an opportunity for dynamic modulation of the phonon sideband of the defect level excitations via strain.
Mechanical strain can be seamlessly applied to TMDC monolayers~\cite{moon_dynamic_2019} which can influence the SOC and result in modulation of optical efficiencies in TMDCs\cite{zollner_strain-tunable_2019}. 
These phonon-assisted effects could be further studied for localized states in TMDCs to understand the contribution of dynamic strain induced modulation of SOC on the transition efficiencies. This would provide a route to controlling an array of defects hosted in TMDC without local engineering of properties at each individual defect site.

\begin{acknowledgments}
This work was supported by the Department of Energy `Photonics at Thermodynamic Limits Energy Frontier Research Center under grant DE-SC0019140. Chakraborty is partially supported by the U.S. Department of Energy, Office of Science, Basic Energy Sciences (BES), Materials Sciences and Engineering Division under FWP ERKCK47 `Understanding and Controlling Entangled and Correlated Quantum States in Confined Solid-state
Systems Created via Atomic Scale Manipulation'.
Ciccarino is partially supported by the Army Research Office MURI (Ab-Initio Solid-State Quantum Materials)
grant number W911NF-18-1-0431.
P.N. is a Moore Inventor Fellow through Grant GBMF8048 from the Gordon and Betty Moore Foundation.
Calculations were performed using
resources from the Department of Defense High Performance Computing
Modernization program. Additional calculations were performed 
using resources of the National Energy Research
Scientific Computing Center, a DOE Office of Science User
Facility, as well as resources at the Research Computing
Group at Harvard University. The authors thank Dr. Bruno Schuler (Swiss Federal Laboratories for Materials Science and Technologies and Lawrence Berkeley Lab) and Dr. John P. Philbin (Harvard University) for useful discussions.

\end{acknowledgments}

%

\end{document}